\documentclass  [11pt] {article}
\title {\bf Cooling-rate effects in a model of (ideal?) glass \vskip5mm}   
\author {{\sc A.~Lipowski}$^{1),2)}$ and {\sc D.~Johnston}$^{1)}$\\
\noalign{\vskip5mm}
$^{^{1)}}${\it Department of Mathematics, Heriot-Watt University},\\ 
{\it EH14 4AS Edinburgh, United Kingdom}\\
$^{2)}${\it Department of Physics, A.~Mickiewicz University,}\\
{\it  61-614 Pozna\'{n}, Poland}\\
\noalign{\vskip5mm}}
\date {}
\newif\ifetykiety

\def\etykieta#1{\ifetykiety \par\marginpar{\centering[#1]} \fi}
\def\bibetykieta#1{\ifetykiety \marginpar{\renewcommand{\baselinestretch}{0.9}
                   \raggedright\small[#1]} \fi}

\newcommand {\SECTION} [2] {\section{#2} \label{#1} \etykieta{#1} \setcounter 
  {equation} {0}}
\newcommand {\SUBSECTION} [2] {\subsection{#2} \label{#1} \etykieta{#1} }

\newcommand {\eq} [1] {(\ref {#1})}
\newcommand {\beq} {\begin {equation}}
\newcommand {\eeq} [1] {\label {#1} \end {equation} \etykieta{#1}}
\newcommand {\beqn} {\begin {eqnarray}}
\newcommand {\eeqn} [1] {\label {#1} \end {eqnarray} \etykieta{#1}}
\catcode `\@=11
\def\@cite#1#2{#1\if@tempswa , #2\fi}
\catcode `\@=12
\newcommand {\cyt} [1] {$^{\mbox {\footnotesize \cite{#1})}}$}

\def\bib#1#2\par{\bibitem{#1} #2 \bibetykieta{#1}}
\newcommand {\fig} [1] {Fig.~\ref{#1}}
\newcounter {fig}
\newenvironment {figure_captions} {\newpage \thispagestyle {empty} \section*
{Figure captions} \begin {list} {\bf Fig.~\arabic{fig}:} {\usecounter{fig}
\settowidth{\labelwidth}{Fig.~999:} }}{\end{list}}
\def\elem#1#2\par{\item#2\label{#1}\etykieta{#1}}

\renewcommand {\baselinestretch} {1.0}
\renewcommand {\cyt} [1] {{\mbox [\cite{#1}]}}
\begin {document}
\maketitle
\begin {abstract}
Using Monte Carlo simulations we study cooling-rate effects in a three-dimensional
Ising model with four-spin interactions.
During coarsening, this model develops growing energy barriers which at low temperature lead to
very slow dynamics.
We show that the characteristic zero-temperature length increases very slowly with the
inverse cooling rate, similarly to the behaviour of ordinary glasses.
For computationally accessible cooling rates the model undergoes an ideal glassy
transition, i.e. the glassy transition for very small cooling rate coincides a thermodynamic
singularity.
We also study cooling of this model with a certain fraction of spins fixed.
Due to such heterogeneous crystallization seeds the final state strongly depends on the
cooling rate.
Only for sufficiently fast cooling rate does the system end up in a glassy state while slow
cooling inevitably leads to a crystal phase.
\end {abstract}
\SECTION{s1}{Introduction}
Although intensively studied for several decades~\cyt{GOTZE}, glasses are still not fully
understood due to their very complicated structure.
However, gradual progress can be clearly observed.
Recently, very interesting theoretical results were obtained concerning for example, aging in
some models with glassy dynamics~\cyt{KURCHAN}.
From the theoretical point of view, one of the problems is the lack of sufficiently
simple models of glasses.
Although recently important progress has been made, the most realistic off-lattice models still
constitute an enormous computational challenge~\cyt{KOB}.
A possible alternative might be lattice models.
Even when existence of the lattice structure is questionable, such simplified
models sometimes do provide a satisfactory description of a macroscopic system.
A prime example is that the critical point of certain binary alloys is in the universality
class of the three-dimensional ferromagnetic Ising model~\cyt{ALS}.
However, an Ising model with only ferromagnetic interactions is not a good candidate for a
model of glasses, since its relatively fast dynamics cannot trap the system in the disordered
(glassy) phase and the system quickly reaches the low-temperature (crystal) phase.
The simplest way to slow down the dynamics is to introduce randomness into the Hamiltonian of
the model~\cyt{YOUNGBINDER}.
However, glasses under certain experimental conditions might be transformed into
translationally invariant crystals, and it is unlikely that random Hamiltonians lead to
translationally invariant solutions.
This suggest one should look for translationally invariant Hamiltonians with a glassy phase
resulting exclusively from the dynamics of the model and not from built-in randomness.

In random systems slow dynamics is mainly due to energy barriers~\cyt{FISHER}.
Is it possible to generate energy barriers in non-random models?
A positive answer to this question was given some years ago by Shore et al.~\cyt{SHORE,SS}.
who showed that in the three-dimensional Ising model with nearest- and frustrating
next-nearest-neighbour interactions (SS model) there exist energy barriers, which diverge with the
size of correlated regions.
These barriers are due to the fact that the energy of an excitation in this model depends not
only on the area of its boundary (like in the ordinary Ising model) but also on the total length of
edges of this boundary.
At sufficiently low temperature, due to these barriers, the dynamics of the model is able to
trap the system in the disordered phase, which we can tentatively identify as a glassy phase.
The trapping mechanism is effective only for temperatures below the corner-rounding transition
$T_{{\rm cr}}$.
For $T>T_{{\rm cr}}$ the trapping mechanism is not effective, fast (ordinary) dynamics is
restored, and the system quickly evolves toward the low-temperature phase.
However, Shore et al.~ argued that their model is not yet a satisfactory model of glasses because these
barriers vanish at the corner-rounding transition, which in turn implies unrealistically fast
increase of zero-temperature characteristic length $l_0$ with the inverse cooling rate $r^{-1}$.
It would thus be interesting to look for some other non-random models which generate energy
barriers and which, hopefully, would be free of this deficiency.

Recently, it has been shown that the three-dimensional Ising model with plaquette interactions also
generates diverging energy barriers which slow down the low-temperature
dynamics~\cyt{LIPO,LIPDES}.
The energy barriers in this model appear due to the same mechanism as in the SS model.
However, in a number of respects the behaviour of this model is quite different from the SS
model.
Firstly, there exists a temperature $T_{{\rm g}}$ (later identified as a glassy transition
temperature) which separate two regimes:
For $T>T_{{\rm g}}$ the random quench, after short transient, reaches the liquid phase in
which it is seems to be stable, at least during computationally accessible time scale.
For $T<T_{{\rm g}}$ the random quench evolve toward the low-temperature phase, but due to the
above mentioned energy barriers, it gets trapped in the glassy phase.
However, the crystal sample undergoes a transition into the liquid phase at the
temperature which is considerably higher than the glassy transition $T_{{\rm g}}$.
It means that in a certain temperature range, due to very strong metastability, the system 
remains either in a crystal or liquid phase, depending on the initial configuration.
In addition, numerical calculations suggest\cyt{LIPDES} that in this model energy barriers exist
even for temperatures $T>T_{{\rm g}}$, and thus this model might be free of
the cooling-rate anomaly of the SS model.

The objective of the present paper is to examine the behaviour of this model under cooling.
We show that the characteristic zero-temperature length indeed increases much more slowly
with the inverse cooling rate than in the SS model and it is probable that this increase is
logarithmic $l_0\sim - \ln r$, as is expected for ordinary glasses.

However, for extremely slow cooling rates such an increase of $l_0$ in our model is unlikely
to hold.
We observed~\cyt{LIPDES}, that although very strong, metastability in this model is only a
quantitative effect and for a sufficiently large system size a droplet nucleation mechanism
should be effective.
It means that for temperature below the critical temperature (which is determined from the crossing
point of the free energies of the liquid and crystal phases) the model prepared in the liquid state
should collapse onto the crystal (or glassy) phase within a finite time.
However, the estimated size of critical droplets is rather large which suggests~\cyt{LIPDES} that
this finite time is also large.
Thus, cooling rates needed to observe such a collapse are presumably computationally
inaccessible.

Some time ago Anderson proposed~\cyt{ANDERSON} that the glassy transition, which is a kinetic
phenomenon, might be related in the limit of small cooling rate with a certain thermodynamic
transition.
The results of the present paper show that the Ising model with a plaquette interactions
provides an interesting realization of this idea: the peak in the specific heat of the liquid
occurs exactly at the temperature where the internal energy jumps under very slow cooling.

Anderson's idea has had a rather limited experimental support~\cyt{JACKLE}.
The main problem is that under slow~\cyt{MATERIAL} cooling real liquids do not get trapped in the
glassy phase but instead crystalize.
The reason for that is that when liquid is cooled below the melting point it becomes metastable and
within a finite time due to heterogeneous or homogeneous crystal nucleation~\cyt{ELLIOT} it
crystalizes.
Only under sufficiently fast cooling can the crystal nucleation be avoided and the liquid
be trapped in the glassy state.
In this context, the model with plaquette interactions corresponds to an almost ideal 
liquid with an extremely large lifetime of a metastable state.
Although such strong metastability allows us to examine the interesting regime of slow
cooling, it is also inhibits the crystalization of supercooled liquid.
To study the competition of crystallization and glass formation within computationally
accessible times we enhanced the former effect by fixing a certain fraction of spins.
We observe that due to such heterogeneous crystallization seeds the final state of the system
indeed strongly depends on the cooling rate.
Namely, only for sufficiently fast cooling can the system beat the crystalization trap and
end up in a glassy state.
When the cooling is slow, similarly to real liquids, the system crystalizes.
Thus, in agreement with experiments, the glassy transition appears to be a kinetic phenomenon with
the cooling rate determining the final state of the system.
These results appear to indicate that the three-dimensional Ising model with plaquette
interactions is a very promising candidate for a lattice model of glasses.

In section 2 we introduce the model and briefly describe its already reported, rather unusual,
properties.
In section 3 we study the behaviour of our model under continuous cooling.
The analysis of the results in presence of crystallization seeds is done in section 4.
Final discussion of our results, including the relation with the ideal glassy transition, is
presented in section 5.
In this section we also argue why our model, being in some sense fine-tuned, might shed some light
on  the apparent robustness of glasses.
\SECTION{s2}{Model and its basic properties}
In the present paper we study the three-dimensional Ising model with a four-spin interaction.
Models with multi-spin interactions have frequently been used, for example, in the context of
random surfaces~\cyt{CAPPI} or lattice field theory~\cyt{SAVVIDY,SAV,CIRILLO}.
There are also some reports of glassy behaviour in such systems~\cyt{MULTI,MEZ}.
Our model is defined by the following Hamiltonian
\beq
H = -\sum S_iS_jS_kS_l,
\eeq {1}
where the summation is over elementary plaquettes of the cubic lattice and $S_i=\pm 1$.
This model was recently studied in the context of lattice field theory~\cyt{BAIG}.
Moreover, the glassy behaviour was studied for the random version of model~\eq{1}~\cyt{ALVAREZ}.
Clearly, a ferromagnetic configuration minimizes the Hamiltonian~\eq{1}.
It is also easy to realize that flipping coplanar spins does not change the energy.
Thus any configuration obtained from the ferromagnetic configuration by flipping coplanar 
spins is also a ground-state configuration.
Also any combination of such coplanar flippings (even for crossing planes) does not increase
the energy.
Simple analysis along these lines shows that for the model on the lattice of the linear size
$L$ the degeneracy of the ground state is equal to $2^{3L}$.
Although the ground state of this model is strongly degenerate, its ground-state entropy is zero.

All the results reported in this paper were obtained using a standard Monte Carlo method
with random sequential update using Metropolis algorithm~\cyt{BINDER}.
Some other details concerning these simulations can be found elsewhere~\cyt{LIPDES}.
\SUBSECTION{ss1}{Thermodynamics and metastability}
Upon heating  an arbitrary ground-state configuration, the model undergoes a sharp transition
at the temperature $T\sim 3.9$, where we have put the Bolzmann constant ${\rm
k}_{{\rm B}}$ to unity~\cyt{LIPO,BAIG}.
This transition is accompanied by a pronounced peak in the specific heat.
The system sizes $L=24$ and 40 used in these simulations were rather large  and the location
of this peak is almost independent on $L$~\cyt{LIPDES}.
These results suggest that the model undergoes a thermodynamic transition around $T=3.9$.

It is thus surprising that upon cooling a high-temperature (liquid) sample the model does
not undergo any change at $T=3.9$.
Instead, it is only when cooled below $T=T_{{\rm g}}\sim 3.4$ that the liquid looses its stability
and
evolves toward the low-temperature phase.
We observed that for $3.4<T<3.9$ it is virtually impossible to direct the evolution
of a liquid sample toward a low-temperature phase.
The transition at $T=T_{{\rm g}}$ is also accompanied by a peak in the specific heat and
for the examined system sizes, $L=24$ and 40, the location of this peak is also almost independent
of $L$~\cyt{LIPDES}.
The behaviour of the specific heat in the vicinity of $T_{{\rm g}}$ is shown in~\fig{heat}.

Using thermodynamic integration we calculated the free energy of both
liquid and crystal phases of the model~\cyt{LIPDES} which show that the crossing point of these
free energies is around $T=3.6$.
However, no changes were observed at that temperature during the heating or cooling.
The above described thermodynamic properties suggests that at computationally
accessible time scale, model~\eq{1} undergoes two transitions
depending whether the system is being cooled or heated.
These transitions seem to screen the 'true' first-order thermodynamic transition which presumably
takes place around $T=3.6$ i.e., at the crossing point of the free energies.

Such behaviour of model~\eq{1} resembles hysteresis and metastability effects, which frequently
occur in ordinary first-order transitions.
However, it is believed~\cyt{METAS} that for short-range interacting systems such effects are only
quantitative and longer simulation time decreases the hysteresis range and eventually pinpoint the
temperature of the first-order transition.
On the other hand, our simulations~\cyt{LIPDES} suggest that in model~\eq{1} squeezing
the hysteresis into the temperature range smaller than $(3.4,3.9)$ is almost
impossible~\cyt{LIFETIME}.
The only way to overcome the very strong metastability of model~\eq{1} is to start simulations
from inhomogeneous initial configurations, i.e., containing both phases of the system.
Indeed one observes~\cyt{LIPDES} that evolution of such a system depends on whether temperature is
above or below the expected thermodynamic transition $T=T_{{\rm c}}=3.6$~\cyt{COMMETAS}.

In the last section we suggest that the metastability of model~\eq{1}, which is probably strongly
related with its glassy properties, is caused by energy and entropy barriers.
However, more precise understanding of the mechanism generating such a strong metastability is 
clearly desirable.
\SUBSECTION{ss2}{Domain coarsening and energy barriers}
When cooled below its critical point, a macroscopic system undergoes the interesting phenomenon
of domain coarsening~\cyt{BRAY}.
Various theoretical and numerical techniques predict that for systems with a scalar order
parameter and nonconserved dynamics, as is the case here, the characteristic length $l$ (which
approximately corresponds to the average size of domains) should increase with time $t$ as 
\beq
l\sim t^{1/2}.
\eeq{1a}
However, as shown by Shore et al.~\cyt{SHORE}, for certain models of this kind the increase of
$l$ can be much slower.
Namely, they have shown that for the SS model and sufficiently low temperature the characteristic 
length $l$ increases only logarithmically in time ($l\sim {\rm ln}t$)~\cyt{RAO}.
Such a slow increase is compatible with the widely accepted conception of glasses.

Recently we have noted that for model~\eq{1} the characteristic length also increases very
slowly in time, presumably also logarithmically.
An additional support to the fact that $l$ might increase in the same way as in the SS model comes
from the fact that both models at low temperature generate energy barriers in the same manner.

To see how these barriers arise in model~\eq{1}, let us consider first its two-dimensional
(square lattice) version.
In particular, let us consider a square domain of '--' spins of linear size $M$ surrounded by '+'
spins (see~\fig{config}a).
Elementary counting~\cyt{LIPO} immediately shows that the energy excess of such a domain is
independent of its size $M$ and depends only on the number of corners in this domain (i.e.,
four).
Such a dependence of energy of excitation on its size should be contrasted with the ordinary
(two-spin) Ising model, where this excess energy is proportional to the perimeter of the
excitation (i.e., $4M$). 
Next, let us observe that to remove such an excitation the system has to flip some of
the '--' spins, but this will inevitably increase the number of corners in the resulting domain
and thus the energy (see~\fig{config}b).
This argument easily generalizes to three dimensions: for the cubic-like domain the excess
energy is proportional to the total length of boundary edges (i.e., $12M$).
Again this is in contrast to the ordinary Ising model, where the excess energy is proportional to
the total area of the boundary (i.e., $6M^2$).
Similarly to the two-dimensional case, to remove such an excitation the system has to climb some
energy barriers, which this time will increase linearly with $M$.
At low temperature such barriers make the process of removing such excitations extremely
slow.
Similar arguments were more thoroughly elaborated for the SS model~\cyt{SHORE}.

However, to show that such barriers are relevant in the process of coarsening one has to show
that the system spontaneously generates such cubic configurations.
Snapshot configurations for the SS model clearly show~\cyt{SHORE} that the system indeed
generates such configurations.
However, due to strong degeneracy of the ground state, the situation is more
complicated for model~\eq{1}.
Firstly, let us notice that low-energy domain walls, as between '+' and '--' domains, are not the
only possibility.
For example, a cubic-like antiferromagnetic domain (antiferromagnetic configuration is also one
of the ground states) surrounded by '+' spins, as in~\fig{config}c, has an excess energy
proportional to the area of the boundary (i.e., like in the ordinary Ising model).
There are also some other ground states, for which energy of domain walls in some sense
interpolates between these low- and high-energy examples.
The extent to which these different domains will appear in the late-time configurations is
determined by a very complicated dynamic process.
In general, however, for high-energy domain walls the energy barriers are much smaller or even
non-existent and we expect that they will be relatively quickly eliminated and the late-time
evolution will be dominated by dynamics of low-energy (and high-barrier) domains.
To some extent this is confirmed in~\fig{f1} which shows a zero-temperature snapshot configuration
obtained during a cooling process, which is described in more details in the next section.
Although we would need the whole three-dimensional structure to draw domain boundaries, we can see
that indeed a lot of relatively large cubic-like (flat) ferromagnetic domains exist and they are
presumably the principal reason for the slow dynamics of model~\eq{1}.
This argument will be also used in the next section to relate the energy excess and the
characteristic length.

To summarize this subsection, our simulations~\cyt{LIPDES} suggest that the low-temperature (i.e.,
at $T<T_{{\rm g}}$) coarsening in model~\eq{1} is very slow, which is presumably related with 
energy barriers which the model can spontaneously generate during such a process.
An independent confirmation of the model's slow dynamics is presented in the next section.
\SECTION{s3}{Cooling}
A glassy transition is essentially a kinetic phenomenon, which appears when a physical system is
being cooled.
Usually one prepares the system at a certain temperature above the glassy transition and then lowers the
temperature e.g., at a constant cooling rate $r=\frac{dT}{dt}$.
One of the important quantities describing this process is a zero-temperature characteristic length
$l_0$ which can be regarded as an average size of domains at the end of the cooling process (i.e., at
$T=0$).
Of course the slower the cooling the larger the characteristic length $l_0$, since the system has more
time to built some local order.
However, for glasses the growth of domains is very slow.
More precisely, on phenomenological grounds, one expects~\cyt{SHORE} that in glasses $l_0$ increases only
logarithmically with the inverse cooling rate, namely
\beq
l_0\sim {\rm ln}(1/r).
\eeq{2}
Such a slow growth of $l_0$ might be contrasted with a much faster one, 
\beq
l_0\sim r^{-1/2},
\eeq{3}
which appears in an ordinary Ising model~\cyt{CORNELL}.
Actually, it is conjectured that the exponents entering asymptotic expressions ~\eq{1a} and~\eq{3}
are the same also for other types of dynamics.
With this conjecture, the relation ~\eq{2} is simply a consequence of the fact that for glasses the
characteristic length $l$ is expected to grow logarithmically in time.

However, from the fact that the model has a slow low-temperature dynamics does not follow that
$l_0$ also slowly increases as function of inverse cooling rate.
This is clearly the case of the SS model: when prepared at a temperature above the critical temperature
and submitted to some cooling, the model inevitably has to pass through the fast-dynamics
temperature range.
For the small cooling rate the growth of order in this temperature range is dominant and thus $l_0\sim
r^{-1/2}$ follows.
Such a rapid increase of $l_0$ is the main reason why SS is not yet a satisfactory model of glasses.
In the following we present some numerical data which show that in model~\eq{1} $l_0$ increases
much slower than in the SS model and we believe that the growth might be even consistent
with~\eq{2}.

We simulated model ~\eq{1} under continuous cooling with a constant cooling rate $r$ and initial
temperature $T_0=4.2$ ($T_0>T_{{\rm g}}$).
This means that temperature as a function of time is given by $T=T_0-rt$.
The temperature dependence of internal energy is shown in~\fig{internal}.
We performed calculations for several system sizes $L$ in order to ensure that $L$ was sufficiently
large.
For example for $r=0.02$ the system size $L=30$ is sufficient to obtain size-independent results but for
$r=0.00002$ we had to take $L=70$.
One can see that although $r$ decreases by three decades, the zero-temperature energy $U_0(r)$ very 
slowly approaches the ground-state energy $U_{{\rm gs}}=-3$.
Such a behaviour provides a qualitative confirmation of the glassy dynamics of our model.
For a quantitative comparison we have to relate the excess energy $\delta U(r)=U_0(r)-U_{{\rm gs}}$ with
the characteristic length $l_0$.
Although it is not a rigorously established relation, one can assume that these quantities are related in
the following way~\cyt{SHORE}
\beq
\delta U(r) \sim \frac{1}{l_0}.
\eeq{4}
To find how $l_0$ increases with the inverse cooling rate, we plot $\delta U(r)$ as a function of
$r$ in the double-logarithmic scale and the graph is shown in~\fig{lograte}.
From this plot one can infer that approximately $\delta U(r)\sim r^{0.2}$ which, using~\eq{4}, becomes
$l_0\sim r^{-0.2}$.
However, the data in~\fig{lograte} have positive curvature and the asymptotic increase of $l_0$
might be even slower.
In addition to that we want to argue that relation~\eq{4} might not hold for model~\eq{1} and a
modified relation will lead to even slower increase of $l_0$.

First let us briefly review arguments leading to relation~\eq{4}.
Let us consider an ordinary Ising model on the three-dimensional lattice of linear size $L$.
If characteristic length is equal to $l$, then the number of domains in this system scale as
$(L/l)^3$.
Since the energy associated with each domain scales as $l^2$ (i.e., like the area of the surface of 
domains), thus the total excess energy per site in the system scales as $l^2(L/l)^3 /L^3 = 1/l$
and~\eq{4} follows.
However, as we have mentioned in the previous section, model~\eq{1} might generate low-energy
interfaces whose energy scales as $l$.
Repeating the above arguments for such interfaces immediately implies
\beq
\delta U(r) \sim \frac{1}{l^2_0}
\eeq{5}
instead of~\eq{4}.
Although it is difficult to provide convincing arguments, we would like to argue in favour of
relation~\eq{5} rather than~\eq{4}.
Namely, we suggest that at the end of the cooling process (i.e., at $T=0$), the interfaces in the
system will be mainly of low-energy and high-barrier type similar to those shown in~\fig{config}a.
Indeed, as we have already noticed at the end of the previous section, the high-energy interfaces
are those with the lowest (or even zero) energy barriers and thus their removal is likely to be the
fastest process in the course of cooling the system.
Moreover, as one can see in~\fig{f1}, the substantial portion of our system is indeed occupied by
relatively large and ferromagnetic segments and, as we already mentioned, interfaces between such
domains have low energy (and high barriers).

Inverting~\eq{5} we obtain $l_0\sim (\delta U(r))^{1/2}$ and thus the increase of $l_0$ would be
given by half of the slope in~\fig{lograte}.
This would imply that $l_0$ increases like $r^{-0.1}$ or, taking into account the positive
curvature in~\fig{lograte}, even slower.
But even with~\eq{4} rather than~\eq{5} holding, the increase of $l_0$ ($\sim r^{-0.2}$ or slower)
is very slow and equally likely one can expect that the asymptotic increase is only logarithmic
with $r$.
However, much more extensive simulations would be needed to definitely resolve this issue.
\SECTION{3a}{Crystallization versus glass formation}
Although interesting on theoretical grounds, the slow-cooling regime is very difficult to examine
experimentally.
As we already mentioned in the Introduction, this is because under slow cooling liquids have
sufficient amount of time to nucleate 'crystal seeds' which divert the evolution toward crystal
phase.
To beat the crystallization trap and transform the liquid into glass one has to cool the system
sufficiently fast, which sometimes requires a very sophisticated technique~\cyt{ELLIOT}.

As we already mentioned, the free energies of crystal and liquid phases
of model~\eq{1} crosses around $T=T_{{\rm c}}\sim 3.6$ which means that for $T_{{\rm g}}<T<T_{{\rm 
c}}$ liquid is in the metastable state.
On the other hand the present calculations show (see \fig{internal}) that in this temperature
range even under the slowest, computationally accessible, cooling the crystallization never
happens.
In order to study the competition of crystallization and glass formation we have to enhance the
former process.
We did it by fixing a certain fraction of spins in the 'up' state.
Numerical results for cooling of such a system are shown in~\fig{c005}.
One can see that for fast cooling ($r=0.2, 0.002$) liquid becomes trapped in the
(high-energy) glassy state.
On the other hand, slow cooling enables the system to reach the (low-energy) crystal state.
Let us also note that in the limit $r\rightarrow 0$ the jump in the internal energy seems
to converge to $T=3.6$ i.e., the crossing point of the free energies of the crystal and the liquid
phases.
Such a behaviour is consistent with the fact that for $T_{{\rm g}}<T<3.6$ the liquid is
metastable and when sufficiently enhanced, the crystallization might take place.
\SECTION{s4}{Discussion}
The main goal of the present paper was to examine the behaviour of the Ising model with a four-spin
interaction under cooling.
We have shown that the zero-temperature characteristic length increases very slowly as a function
of inverse cooling rate.
Moreover we have shown that when nucleation seeds are introduced and the cooling is slow enough,
the system ends up in the crystal phase.
Thus in agreement with many experiments, the glassy transition becomes a kinetic phenomenon driven
by the cooling rate.
These results, together with the fact that the model possesses a slow coarsening dynamics, is a
very strong indication that model~\eq{1} might capture the essence of the glassy transition in
realistic systems.
In this section we discuss some other implications of our results.\\
\par
\noindent
(i) {\it Ideal glassy transition}\\
Some time ago it was proposed by Anderson that in the limit of vanishing cooling rate the glassy
transition might be related with a certain thermodynamic transition.
This idea has only limited experimental confirmation~\cyt{JACKLE}: the excess entropy of the
supercooled liquid extrapolates to zero at a temperature, which seems to be close to the
temperature of the singularity of the viscosity, when fitted using the so-called
Vogel-Tammann-Fulcher formula.
However, experimental difficulties lead to very ambiguous results and even the very existence of
this hypothetical thermodynamical transition is still not certain.

More generally, Anderson's idea expresses the expectation that changes of dynamical and
thermodynamical properties should be related.
Indeed, there are some reports supporting this statement.
For example, some works on frustrated and/or disordered systems indicate that the appearance of
nonexponentially decaying correlation functions might be induced by a certain
thermodynamical~\cyt{RANDEIRA} or percolative transition~\cyt{CONIGLIO}.
Another example is the frequently referred to in our paper SS model, where the appearance of
the slow-dynamics regime is induced by the corner-rounding transition, which is a well-defined
equilibrium transition~\cyt{WORTIS}.

In our opinion, the comparison of~\fig{heat} and~\fig{internal} shows that model~\eq{1} conforms
Anderson's idea.
The specific heat shown in~\fig{heat} is a thermodynamic quantity.
It was calculated in a (standard) quasi-equilibrium manner: after fixing a temperature we relaxed
the system and then measured the variance of the internal energy.
The sharp peak seen in~\fig{heat} indicates a thermodynamic-like singularity in this model.
On the other hand an almost vertical drop of the internal energy under continuous cooling, shown
in~\fig{internal}, indicates the proper (i.e., dynamic) glassy transition.
One can see in~\fig{internal} that the lower the cooling rate $r$ the sharper the transition.
At first sight one might expect that in the limit $r\rightarrow 0$ the transition becomes
infinitely sharp and coincides with thermodynamic singularities, as e.g., the peak of the specific 
heat.
However, as we already mentioned, our previous results~\cyt{LIPDES} suggest that metastability of
the liquid is only a finite time/size effect and neither the peak nor the internal energy drop can
be made perfectly sharp.
Such a scenario, is in agreement with experimental description of the glassy
transition~\cyt{GOTZE}.\\ 
\par
\noindent
(ii) {\it Mechanism of the glassy transition}\\
Which properties of our model are responsible for the glassy behaviour?
The mechanism which we would like to suggest is based on the presence of both entropy and energy
barriers.
The latter ones were already discussed in this paper and it is likely that they are the source of 
the slow low-temperature dynamics of model~\eq{1} and also of the SS model.
But they are not sufficient for the model to be glassy: in the SS model these barriers vanish below
the critical point and as a result the model orders too quickly.

Measurements of a certain characteristic time suggest that energy barriers in model~\eq{1} exist
even above the glassy transition~\cyt{LIPDES}.
When this property is combined with the ability of the liquid to persist down to $T_{{\rm g}}$, it
leads to a remarkable consequence: when cooled below $T_{{\rm g}}$, the liquid looses stability and
the domain coarsening starts.
But the strong energy barriers, which exist between certain types of domains hampers the process
and trap the system in the glassy phase.
Consequently, there is no fast-dynamics regime in the low-temperature phase and upon cooling the
system orders very slowly.
The second crucial property is thus the ability of the liquid to persist (for some time) in the
metastable state.
Since the liquid state is highly disordered, we expect that it is supported by some entropy
barriers~\cyt{ENTROPY}.
The only argument which we can provide to justify this claim is that these entropy barriers
are presumably related with the strong degeneracy of the ground state of model~\eq{1}, which leads
to many different types of small domains in the liquid phase.
In the SS model the ground state is only double degenerate and apparently such barriers are absent.

We have to emphasize, however, that the very concept of energy and entropy barrier is only
approximate.
At finite temperature it is more appropriate to consider rather free-energy barriers but then the
whole mechanism becomes much more difficult to comprehend.
We hope that further studies will explain properties of this model in a more rigorous way.\\
\par
\noindent
(iii) {\it Model~\eq{1} and robustness of glasses}\\
It is sometimes claimed that virtually every liquid can be transformed into glass, provided that
the cooling rate is fast enough~\cyt{ELLIOT} and any model of glasses should explain such
robustness of glasses.
As we already mentioned, the glassy behaviour in our model results from the presence of both
entropy and energy barriers.
How robust might such properties be in real systems?
Firstly let us consider the energy barriers.
The SS model contains competing nearest- and next-nearest neighbour interactions and model~\eq{1}
contains only multiple interactions.
Since both types of interactions are quite common in real systems we expect that such barriers 
might be a rule rather than an exception.
In our opinion, it is the absence of energy barriers (as in two-spin Ising model) which is unlikely
to exist in real systems.
As for the entropy barriers they are also likely to be common in real systems~\cyt{ENTROPY}.
However, when model~\eq{1} is perturbed, for example, by adding a two-spin interaction, then the
ground state becomes  double degenerate (as the SS model) and entropy barriers will presumably
disappear.
Such a fragile nature of entropy barriers in model~\eq{1} might be a consequence of the
discreteness of the model.
Thus, retaining plaquette interactions only in model~\eq{1} does not mean that we expect that only
systems with such fine-tunned interactions exhibit glassy behaviour.
Such a choice is needed only to generate entropy barriers in lattice models.
In off-lattice models such barriers are likely to be more generic.
\begin {thebibliography} {00}

\bib {GOTZE} W.~G\"{o}tze, in {\it Liquid, Freezing and Glass Transition}, Les Houches Summer
School, ed. J.~P.~Hansen, D.~Levesque and J.~ Zinn-Justin (North-Holland, Amsterdam, 1989).
C.~A.~Angell, Science {\bf 267}, 1924 (1995).
F.~H.~Stillinger, Science~{\bf 267}, 1935 (1995).

\bib{KURCHAN} L.~F.~Cugliandolo and J.~Kurchan, Phys.~Rev.~Lett.~{\bf 71}, 173 (1993).

\bib {KOB} For review see: W.~Kob, in {\it Annual Reviews of Computational Physics}, edited by
D.~Stauffer (World Scientific, Singapore, 1995) vol.~III.

\bib{ALS} J.~Als-Nielsen, in {\it Phase Transitions and Critical Phenomena}, eds.~C.~Domb and
M.~S.~Green (Academic Press, London, 1976).

\bib{YOUNGBINDER} K.~Binder and A.~P.~Young, Rev.~Mod.~Phys.~{\bf 58}, 801 (1986).

\bib{FISHER} D.~S.~Fisher, Phys.~Rev.~Lett.~{\bf 56}, 416 (1986).

\bib {SHORE} J.~D.~Shore, M.~Holzer and J.~P.~Sethna, Phys.~Rev.~B {\bf 46}, 11376 (1992).

\bib {SS} J.~D.~Shore and J.~P.~Sethna, Phys.~Rev.~B {\bf 43}, 3782 (1991).

\bib {LIPO} A.~Lipowski, J.~Phys.~A {\bf 30}, 7365 (1997).

\bib {LIPDES} A.~Lipowski and D.~Johnston, J.~Phys.~A, submitted (cond-mat 9812098).

\bib {ANDERSON} P.~W.~Anderson, in {\it Ill Condensed Matter}, ed.~R.~Balian, R.~Maynard and
G.~Toulouse (North Holland, Amsterdam, 1979).

\bib {JACKLE} J.~J\"{a}ckle, Rep.~Prog.~Phys.~{\bf 49} (1986) 171.

\bib{MATERIAL} The word 'slow' is very much material dependent (see e.g.,~\cyt{ELLIOT}).

\bib{ELLIOT} S.~R.~Elliot, {\it Physics of amorphous materials} (John Wiley \& Sons Inc., New York, 
1990).

\bib{CAPPI} A.~Cappi, P.~Colangelo, G.~Gonella and A.~Maritan, Nucl.~Phys.~B {\bf 370}, 659 (1992)

\bib{SAVVIDY} G.~K.~Savvidy and F.~J.~Wegner, Nucl.~Phys.~B {\bf 413}, 605 (1994).

\bib {SAV} R.~V.~Ambartzumian, G.~S.~Sukiasian, G.~K.~Savvidy and K.~G.~Savvidy, Phys.Lett.~B {\bf 
275}, 99 (1992).

\bib {CIRILLO} E.~N.~M.~Cirillo, G.~Gonella, D.~A.~Johnston and A.~Pelizzola, Phys.~Lett.~A {\bf 
226}, 59 (1997).

\bib{MULTI} J.~Kisker, H.~Rieger and M.~Schreckenberg, J.~Phys.~A {\bf 27}, L853 (1994).

\bib{MEZ} J.~P.~Bouchaud  and M.~M\'{e}zard, J.~Phys.~I (France) {\bf 4}, 1109 (1994).

\bib {BAIG} D.~Espriu, M.~Baig, D.~A.~Johnston and R.~P.~K.~C.~Malmini, J.~Phys.~A {\bf 30}, 405
(1997).

\bib{ALVAREZ} D.~Alvarez, S.~Franz and F.~Ritort, Phys.~Rev.~B {\bf 54}, 9756 (1996).

\bib {BINDER} K.~Binder, in {\it Applications of the Monte Carlo Method in Statistical Physics},
ed.~K.~Binder, (Berlin: Springer, 1984).

\bib {METAS}  O.~Penrose and J.~L.~Lebowitz, in {\it Fluctuation Phenomena}, ed. E.~W.~Montroll
and J.~L.~Lebowitz (Amsterdam: North Holland, 1979).
P.~A.~Rikvold and B.~M.~Gorman, in {\it Annual Review of Computational Physics} vol.I,
ed.~D.~Stauffer (Singapore: World Scientific, 1994).

\bib {LIFETIME} Although we could overcome metastability of this model in direct
simulations, we observed that the droplet nucleation mechanism must be effective.
Namely, when we start the simulations with the thermodynamically unstable phase with a sufficiently
large droplet of a stable phase injected 'artificially' at the beginning then this droplet might
divert the evolution toward the stable phase.
Without such a droplet the unstable phase remains in its state virtually forever.
Moreover, the minimal size of such droplets is rather large which suggests that their spontaneous
nucleation is an extremely unlikely event.

\bib {COMMETAS} With such a choice of the initial configuration the system does not have to
nucleate the stable phase and the evolution is greatly enhanced toward the thermodynamically stable
phase.
 
\bib {BRAY} A.~J.~Bray, Adv.~in Phys.~{\bf 43}, 357 (1994).

\bib{RAO} Additional discussion of the domain growth in the SS model can be found in: M.~Rao and
A.~Chakrabarti, Phys.~Rev.~E {\bf 52}, R13 (1995).

\bib {CORNELL} S.~Cornell, K.~Kaski and R.~Stinchcombe, Phys.~Rev.~B {\bf 45}, 2725 (1992).

\bib {RANDEIRA} M.~Randeira, J.~P.~Sethna and R.~G.~Palmer, Phys.~Rev.~Lett.~{\bf 54}, 1321 (1985).

\bib {CONIGLIO} G.~Franzese and A.~Coniglio, Phys.~Rev.~E {\bf 58}, 2753 (1998).

\bib {WORTIS} C.~Rottman and M.~Wortis, Phys.~Rep.~{\bf 103}, 59 (1984).

\bib {ENTROPY} U.~Mohanty, I.~Oppenheim and C.~H.~Taubes, Science~{\bf 266}, 425 (1994).
T.~R.~Kirkpatrick and P.~G.~Wolynes, Phys.~Rev.~B {\bf 36}, 8552 (1987).
D.~Thirumalai and P.~G.~Wolynes, Phys.~Rev.~A {\bf 40}, 1045 (1989).

\end {thebibliography}
\begin {figure_captions}

\elem {heat} The specific heat $C$ as a function of temperature $T$ calculated from the variance of
the internal energy for $L=24$ ($\diamond$) and $L=40$ ({\large +}).
At each temperature we relaxed the system for $10^3$ Monte Carlo steps and measurement was done
during $10^4$ Monte Carlo steps.

\elem {f1} An example of zero-temperature single-layer configuration obtained during the cooling of
model~\eq{1} at the rate $r=0.0002$ and for the system size $L=50$.
Up and down spins are denoted by $\diamond$ and dots, respectively.

\elem {config} (a) An example of low-energy interface in the two-dimensional version of
model~\eq{1}.
An excess energy (i.e., the number of 'unsatisfied' plaquettes) comes from the four corner
plaquettes.
To remove such a configuration the system is likely to proceed through configurations like those
shown in (b).
The excess energy is higher in this case.
(c) An example of high-energy interface (ferromagnetic and antiferromagnetic states are ground
states of model~\eq{1}).
One can easily see that the excess energy increases linearly with the size of this excitation.
However, to remove this excitation the system does not have to increase its energy (there are no
energy barriers in this case).
The process of removal of such excitations should be much faster and basically such as in the
two-spin Ising model.

\elem{internal} The internal energy $U$ as a function of temperature for (from the top) $r=$ 0.02, 
0.002, 0.0005, 00002, 0.00005 and 0.00002.

\elem{lograte} The excess energy $\delta U(r)$ as a function of $r$ in the double-logarithmic
scale.
The dotted line has slope 0.2.

\elem{c005} The internal energy $U$ as a function of temperature for (from the least to the most
steep) $r=$ 0.02, 0.002, 0.0005, 0.0002, 0.00005, 0.00002, 0.00001 and 0.000002.
Calculations were done for $L=50$ and with 5\% of spins fixed in the 'up' state.
The dotted line corresponds to heating the ferromagnetic state (L=40) without any spins fixed.
The slow-cooling results ($r<0.0005$) in the low-temperature regime are indistinguishable from
those corresponding to heating.

\end {figure_captions}
\end {document}
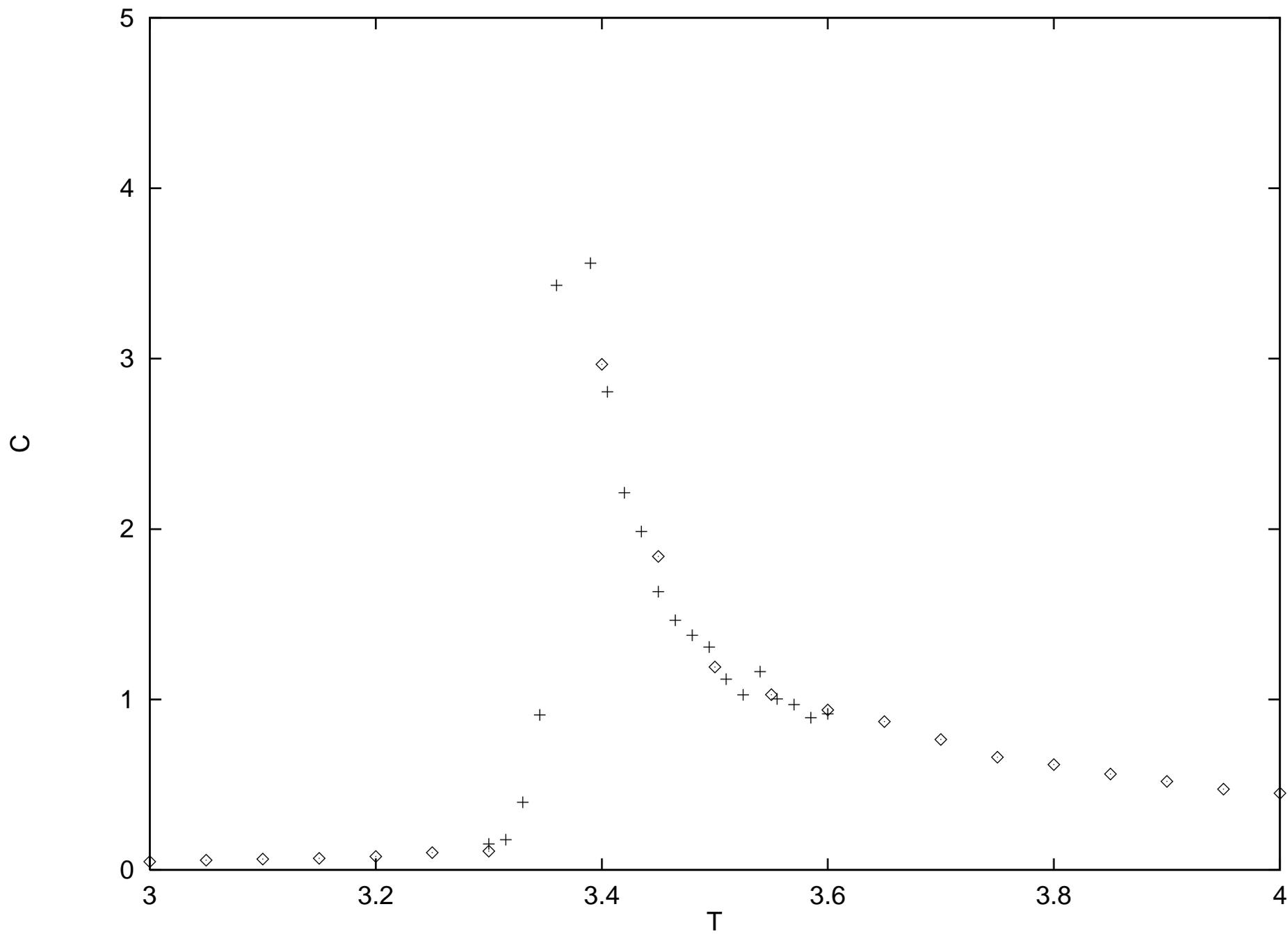
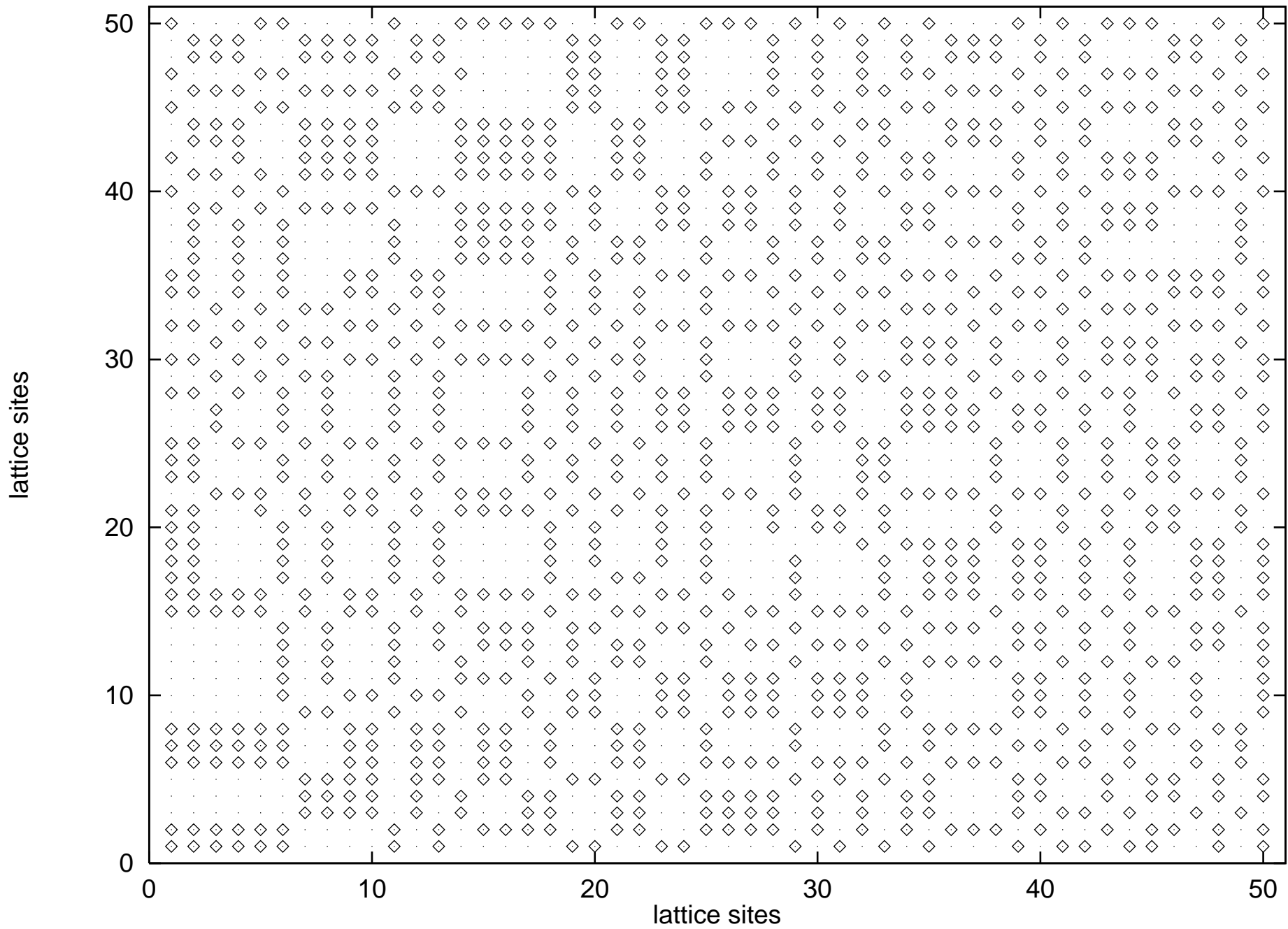
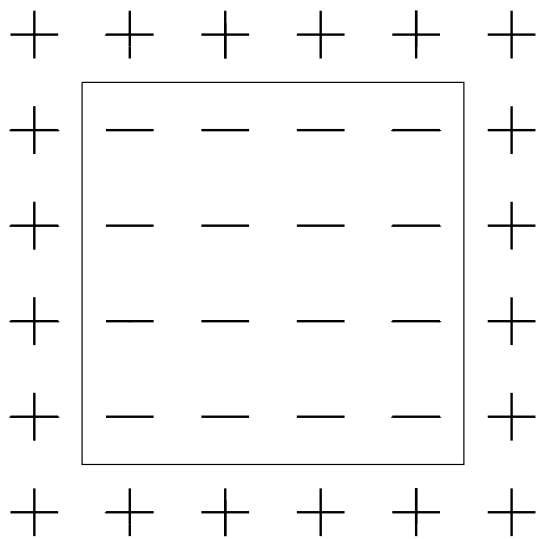
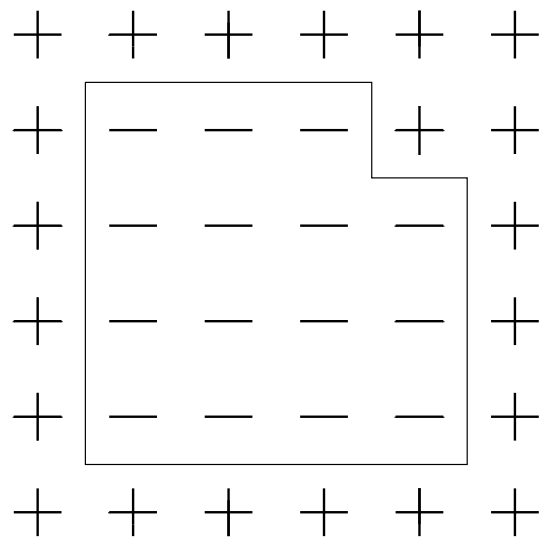
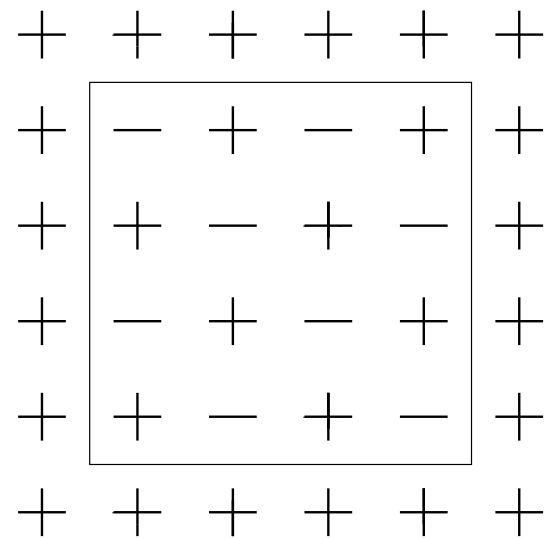
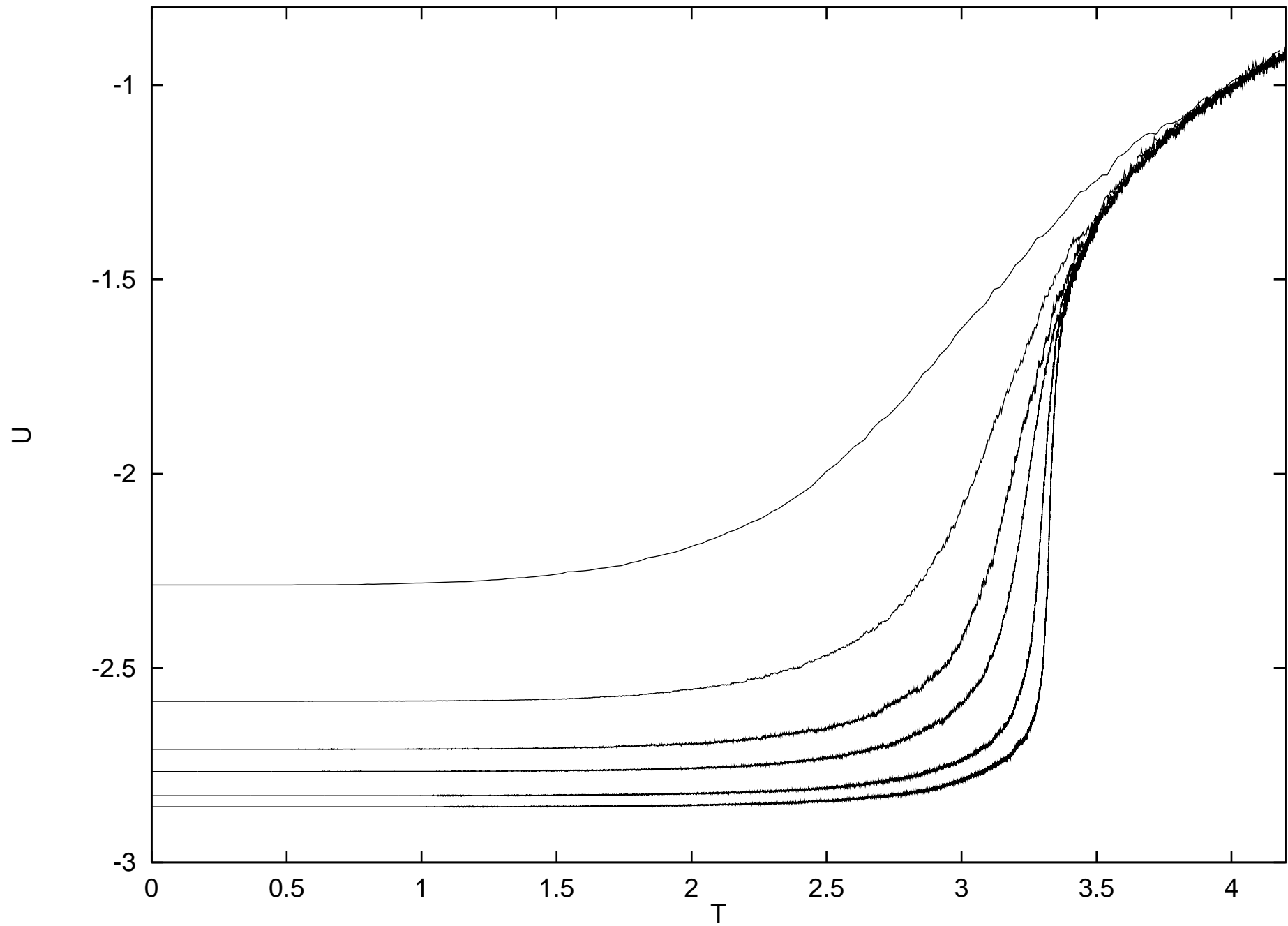
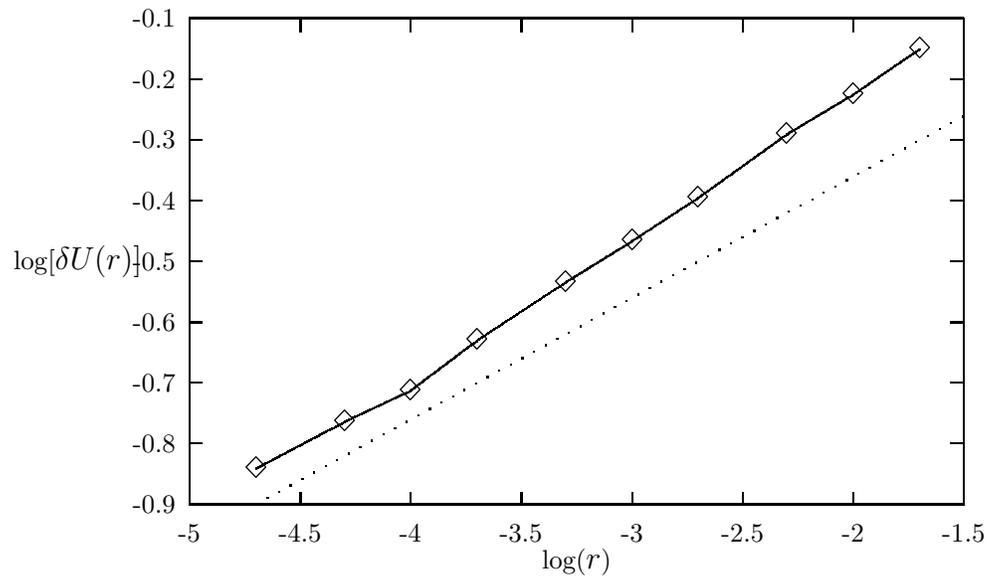
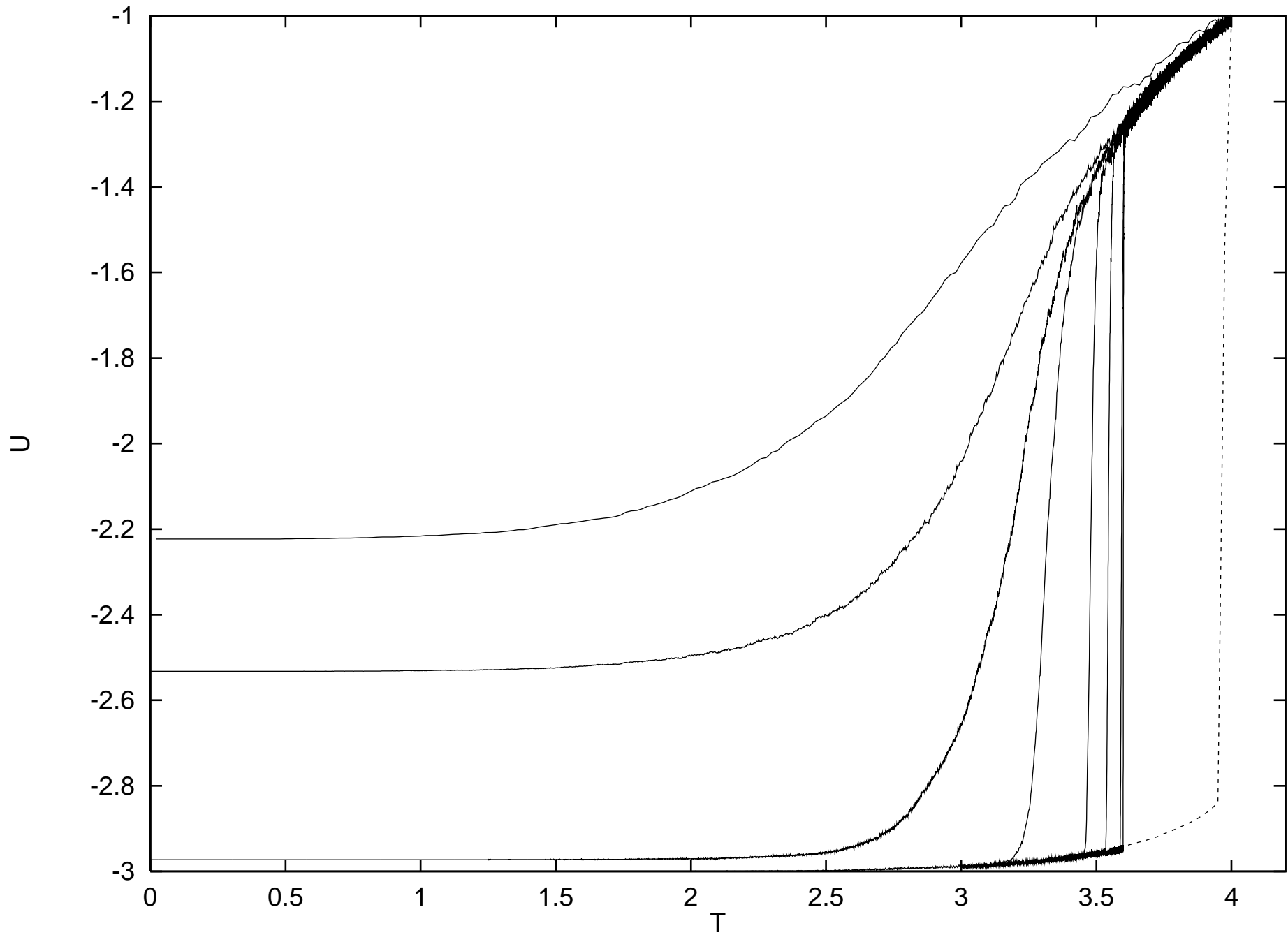
